\documentclass{article}

\usepackage{cite}
\usepackage{amsmath,amssymb,amsfonts}
\usepackage{algorithmic}
\usepackage{graphicx}
\usepackage{textcomp}
\usepackage{xcolor}
\usepackage{fancyhdr}
\def\BibTeX{{\rm B\kern-.05em{\sc i\kern-.025em b}\kern-.08em
    T\kern-.1667em\lower.7ex\hbox{E}\kern-.125emX}}

\date{Department of Physics, University of Trento, Via Sommarive 14, Trento, Italy}
\title{Reservoir Computing Model For Multi-Electrode Electrophysiological Data Analysis}
\author{Ilya Auslender \and Lorenzo Pavesi}
\begin{document}

\maketitle

\begin{abstract}
In this paper we present a computational model which decodes the spatio-temporal data from electro-physiological measurements of neuronal networks and reconstructs the network structure on a macroscopic domain, representing the connectivity between neuronal units. The model is based on reservoir computing network (RCN) approach, where experimental data is used as training and validation data. Consequently, the model can be used to study the functionality of different neuronal cultures and simulate the network response to external stimuli.
\end{abstract}

\section{Introduction}
Electrophysiological study in neuroscience provides a wide-vision of the interplay between cells of different types at different scales \cite{llinas1988intrinsic}. Such studies vary from investigating the function of a single cell up to studying the dynamics of complex systems consisting of a large number of cells \cite{contreras2004electrophysiological}, in the pursuit of obtaining a comprehensive picture of the brain activity. In particular, \textit{in-vitro} studies of neurons give a simplified representation of the structure and functionality of these networks in living organisms \cite{gross1977new,chiappalone2019vitro}. Such approach assists in decomposing the extremely complex structure of living brain into smaller functional blocks. 

As the complexity of the biological system increases, it becomes more and more challenging to analyze or model the behavior in such systems. Numerous models are designed to picture the dynamics behind neuronal activity, starting from single cell models (e.g., Hodgkin–Huxley model \cite{hodgkin1952quantitative}) up to models of large populations \cite{marder2011multiple,gerstner2002spiking}. Various methods focus on the biophysical properties of the cells (e.g. membrane voltage), while others focus on the point-process of information propagation (e.g., spike trains). Some approaches use experimental observations to adapt a model which will be a computational counterpart to the biological system \cite{natschlager2003computer,yada2021physical,dockendorf2009liquid}. Such methods use Machine- or Deep- Learning techniques to train a given model to construct the desired outcome. While for some research questions such approach could be very inefficient and/or computationally expensive, for others it can provide a practical solution to construct a computational tool for various applications.

\begin{figure}[t]
    \centering
    \centerline{\includegraphics[width=3in]{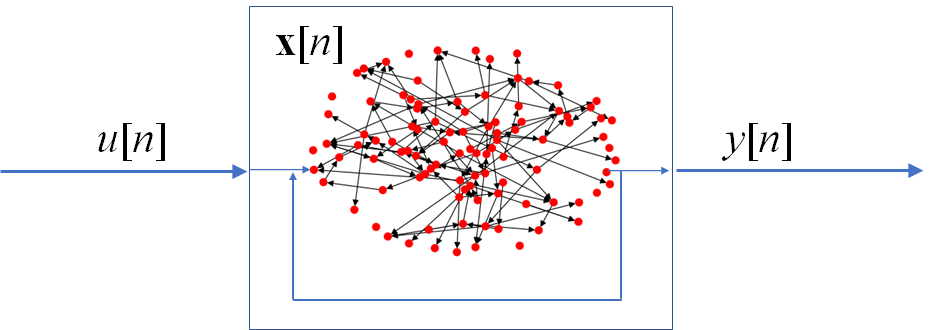}}
    \caption{A general description of reservoir computing network (RCN). Time-sequence $u[n]$ is processed in a higher dimensional space by a non-linear reservoir operator, which updates in recurrent manner, via complex interconnections, a reservoir state $\mathbf{x}[n]$ according to the history of the input. This reservoir state is then transformed in an output $y[n]$.}
    \label{fig:reservoir general}
\end{figure}

We propose in this work a simplified approach for interpreting electrophysiological signals from neuronal networks from which a functional connectivity between neuronal populations is retrieved and an interplay between them can be predicted at a macroscopic level. The model is based on Reservoir computer network (RCN) \cite{lukovsevivcius2009reservoir}, since the information emerged by sampling electrophysiological signals from a cultured neuronal network is obtained from a complex neural circuitry. The complexity of these circuits cannot be easily understood from a standard measurement analysis, and hence they are modeled as nonlinear networks with inner random connections. The general concept of RCN is depicted in Fig. \ref{fig:reservoir general}.

The general form of the RCN dynamics is given by:
\begin{equation}
\mathbf{x}[n] = f\bigl(\mathbf{x}[n-1],u[n]\bigr)
\label{eq: reservoir equation, general}
\end{equation}
and,
\begin{equation}
y[n] = g\bigl(\mathbf{x}[n]\bigr)
\label{eq: output equation general}
\end{equation}
where $u[n],y[n]$ is the input and output signals, respectively, at a discrete time $n$; $\mathbf{x}[n]$ is a reservoir state in a higher dimensional space at a discrete time $n$. $f$ and $g$ are functions.

Using this model we are able to extract a macroscopic graph representing the structure of the culture under test, where each node of the network represents a neural circuit (population of neurons); and the connections (edges) between them represent the weighted interaction between the populations.

\section{Model description}
We consider a multi-site measurement of electrophysiological signals from a neuronal culture, such as 2D microelectrode array (MEA). We seek to represent the tested culture as a network where each node corresponds to one measurement electrode. Each electrode samples the electrophysiological signals from the neuron ensemble (consisting of a few neurons) found in its vicinity. Therefore, each node has to represent a complex neuronal circuit whose dynamics by itself is driven by many interacting neurons. We hence define the domain of the measurement as the \textit{macroscopic domain}, which is described by the network in question; whereas the neuronal structure which is sampled by each node will be referred as the \textit{microscopic domain} (or later as the \textit{reservoir domain}). The data unit which is contained in each of these nodes is a sample of the electrophysiological signals expressed in the instantaneous spike-rate measured in a specified time window. By “data unit” we refer to a set of data sampled at the network nodes in a definite time window, which contains information on the status of the network, with a memory on the previous time steps, and the ability to predict the next step accordingly. The time window is determined by a characteristic rate of the network, which can be obtained, for example, by analyzing the inter-spike interval (ISI) histogram \cite{pasquale2010self}. This unit of time is dependent on many properties of the network such as neuron density in the culture, age of the culture and other \cite{chiappalone2006dissociated}, and it characterizes the signal integration time of each node.

Let us represent the macro-domain state of the network at each time step $n = 1,2,3...$ with a vector $\mathbf{y}$[n], where each component of the vector describes the state of a single node, i.e., $\mathbf{y}[n]$ is the signal representation of each electrode at time $n$. The purpose of this work is to find a time propagation operator $\mathcal{\hat{O}}$, such that:
\begin{equation}
    \mathbf{y}[n+1] = \mathcal{\hat{O}} \Bigl\{\mathbf{y}[n]\Bigr\}\
    \label{Eq: Operator}
\end{equation}
where the operator $\mathcal{\hat{O}}$, which is likely to be non-linear due to the nature of neuronal networks, should describe as closely as possible the experimental observation in the electrophysiological measurements, i.e., we aim to fit a model to an observation which can mimic or predict the spatio-temporal patterns of the neuronal activity in the culture under test.

We then consider the fact that each node in the macro-domain network represents a complex neuronal signal-processing-unit. It arises from the fact that typically every measurement site is surrounded by neurons which may be as many as a dozen. The morphology and functionality of each of these micro-circuits embedded in each node of the macro-domain network cannot be easily obtained from the electrophysiological measurements of standard recording systems. Also modeling of such neuronal structures is not an easy task and has been studied for decades, with numerous models for different scales of dimensions and time \cite{marder2011multiple,gerstner2002spiking}. No strong consensus however is achieved about the authenticity of these models. 

\begin{figure}[t]
    \centering
    \centerline{\includegraphics[width=3.5in]{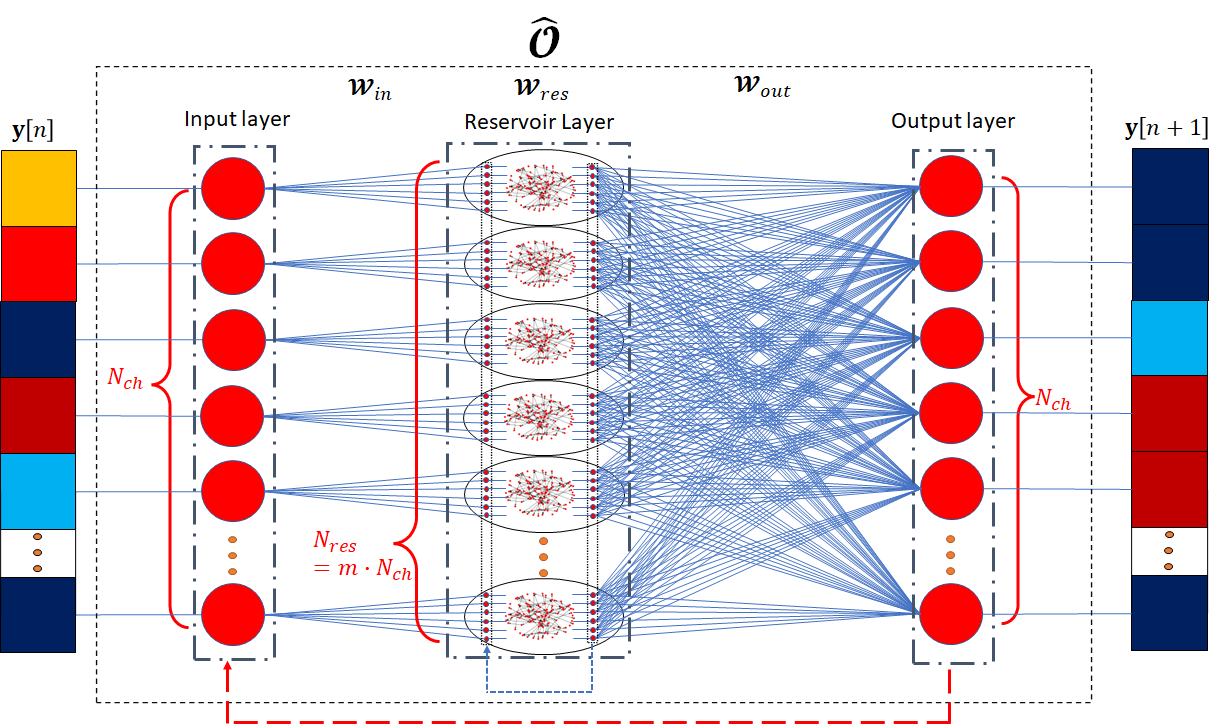}}
    \caption{A scheme of data processing for each time step of the data sequence $\mathbf{y}[n],\mathbf{y}[n+1], \mathbf{y}[n+2] \ldots$ At each time step $n$ a state $\mathbf{y}[n]$, representing the instantaneous activity of the macro-domain network (here encoded in color-scale), is processed by three layers: input, reservoir and output (detailed below) and eventually transformed to the next state of the network, $\mathbf{y}[n+1]$. The whole process is described by the operator $\hat{\mathcal{O}}$ as was defined in \eqref{Eq: Operator}.}
    \label{fig:ANN}
\end{figure}

Hence our approach is to represent each measurement node (electrode) as a gate to a particular neuronal circuit (reservoir), where the signals measured at each node are an outcome of a complex operation involving each circuit and the whole network. We therefore propose the artificial neural network (ANN) structure depicted in Fig. \ref{fig:ANN}. This structure represents a simplification of the neuronal dynamics, where each of the neuronal circuits is a black box, whose morphology and functionality are not known but assumed to be reasonably random. 
As seen in Fig. \ref{fig:ANN}, we assume that the signal sampled at each node is an input to and an output from a higher dimensional domain with a specific connectivity and functionality. At the input of the ANN, the signal at each node is transformed to a corresponding reservoir-state (in the reservoir domain) by a set of uncoupled weighted connections (Input layer). Each micro-reservoir, associated with a node in the macro-domain, represents a micro-neuronal circuit embedded at each of the measurement sites, and has inner interconnections which represent the connectivity of the micro-circuits (Reservoir Layer). Each such circuit performs a nonlinear transformation, creating an updated reservoir state, which on one side is stored as a memory to be integrated to the next time steps, and on the other side is used to form the next state of the macro-domain network by weighting and coupling all the micro-reservoir states (Output Layer); then the whole process repeats cyclically. This kind of recurrent network is known as \textit{Reservoir Computing Network} (RCN) and has been widely studied.

\section{Model Design}
\subsection{Domains and Dimensions}
As mentioned above, the model distinguishes between two domains: The macro-domain which refers to the experimental observations, represented by the corresponding network; and the micro- (or reservoir) domain which refers to the neuronal units embedded in each of the macro network nodes, with no experimental data. We denote by $N_{ch}$ the dimension of the macro-network which in fact represents the number of nodes in the network, where each node is directly associated with an electrode (or a channel) in the experimental measurement. $N_{res}$ is the dimension of the reservoir. 

Assuming that the neurons are uniformly distributed in the culture, we appoint a fixed number of connections between each node and the corresponding micro-circuit, such that for each node of the network there is one micro-reservoir (see Fig. \ref{fig:ANN}):
\begin{equation}
  N_{res} = mN_{ch}
  \label{eq: dimensions}
\end{equation}
where $m$ is an integer number. It follows that each $m$ components in the vector space of the reservoir domain correspond to one node in the macro domain. In fact, we may associate $m$ with a relative size of each micro-circuit.
\subsection{Input Layer}
The input layer refers to the stage between the macro domain and the reservoir one. Here we assume that the data at each of the nodes is a linear transformation of the corresponding input state to the reservoir, such that each component in the macro-domain transforms directly to corresponding $m$ inputs of $N_{res}$ components in the reservoir domain, and refer to a single micro-circuit. This is done with the following transformation:
\begin{equation}
\mathbf{x}_{in}[n] = \mathcal{W}_{in}\mathbf{y}[n]
\label{eq: input}
\end{equation}
where $\mathbf{y}\in \mathbb{R}^{N_{ch} \times 1}$ is the vector representing the state of the network nodes. $\mathbf{x}_{in}\in \mathbb{R}^{N_{res} \times 1}$ is the corresponding vector in the reservoir domain. Therefore, $\mathcal{W}_{in}\in \mathbb{R}^{N_{res} \times N_{ch}}$ is a linear transformation. Since $\mathcal{W}_{in}$ maps each node to a corresponding micro-reservoir, it is represented by the following matrix:
\begin{equation}
\small{
 \mathcal{W}_{in} = 
\begin{bmatrix}
\left (
\mathbf{w}_{in}^{(1)}
\right )

&  0 & 0
& \cdots & 0\\ 
    0 & 
    \left( \mathbf{w}_{in}^{(2)} \right ) & 0
    & \cdots &  0 \\
    0 & 0 & \left( \mathbf{w}_{in}^{(3)} \right ) & & 0\\ 
    \vdots & \vdots & & \ddots & \\
    
    0 & 0 & 0 &  & \left (\mathbf{w}_{in}^{{(N_{ch})}} \right )
\end{bmatrix}
}
\label{eq: W_in}
\end{equation}
where each $\mathbf{w}_{in}^{(i)}\in \mathbb{R}^{m \times 1}$, $i=1,2,...,N_{ch}$ is a vector with random weights taken from a \textit{Normal distribution} (peaked at 0), normalized such that $\|\mathbf{w}_{in}^{(i)}\|^2=1$, which can also be expressed as:
\begin{equation}
   \mathcal{W}_{in}^T \mathcal{W}_{in} = \mathcal{I}_{N_{ch}}
   \label{eq: W_in constraint}
\end{equation}
where $\mathcal{W}_{in}^T$ is the transposed input matrix and $\mathcal{I}_{N_{ch}}$ is the unit matrix of order $N_{ch}$.

\subsection{Reservoir Layer}
The reservoir layer contains $N_{ch}$ independent micro-circuits with $m$ nodes each (total $N_{res}$ nodes). Each such circuit models the neuronal circuit around each electrode. This layer has two main functionalities: 1. nonlinear time-operator. 2. Reservoir state integrator.
In particular we consider the following dynamics for the reservoir network:
\begin{equation}
\mathbf{x}[n] = \mathbf{f}_{NL} \Bigl(\hat{\mathcal{S}}\cdot\left( \mathbf{x}_{in}[n]+\alpha\mathcal{W}_{res}\mathbf{x}[n-1] \right)\Bigr)
\label{eq: reservoir equation non-linear}
\end{equation}
where $\mathbf{x}[n]$ is the reservoir state obtained at time step $n$, from the combination of the input state $\mathbf{x}_{in}[n]$ (given by \eqref{eq: input}) and an inner transformation of the reservoir state at time step $n-1$, $\mathbf{x}[n-1]$. This discrete differential relation provides cumulative data at each time step and carries the temporal memory on the activity of the network. $\hat{\mathcal{S}}$ is a diagonal matrix containing normally-distributed synaptic strengths on its diagonal, expressing the variance of the synaptic nonlinear response of the micro-reservoirs. $\mathbf{f}_{NL}$ is the nonlinear function. Typical functions that are used in this approach are tanh or sigmoid, which have the saturation property and prevent the reservoir from exploding. In this work we tested a few nonlinear functions similar to the mentioned above. $\mathcal{W}_{res}\in \mathbb{R}^{N_{res} \times N_{res}}$ is a matrix, which performs an inner map (i.e. from and back to the reservoir domain) of the reservoir state in previous step to a new state; and it represents the inner connections within each of micro-reservoirs. We assume that $\mathcal{W}_{res}$ is a norm-preserving linear map, i.e. conserving the energy of the state. Therefore, we represent this transformation by an orthogonal matrix with normally distributed random weights, with zero mean. In addition, we do not allow the coupling between the different micro-reservoirs at this point, hence we represent this matrix in the following block-diagonal form:
\begin{equation}
 \mathcal{W}_{res} = 
\begin{bmatrix}
\left (
\mathbf{W}_{res}^{(1)}
\right )
& 0 & \cdots & 0\\
0 & 
\left ( \mathbf{W}_{res}^{(2)} \right )
 & \cdots & 0\\
\vdots & \vdots &  \ddots &  \\
0 & 0 & &
\left ( \mathbf{W}_{res}^{(N_{ch})} \right )
\end{bmatrix}
\label{eq: reservoir matrix}
\end{equation}
where each $\mathbf{W}_{res}^{(i)}\in \mathbb{R}^{m \times m}, i=1,2...N_{ch}$ is a random-orthogonal matrix. Note that each block acts on its corresponding micro-reservoir state. Next we define $0< \alpha < 1$ which is the memory parameter. It expresses the temporal memory strength, i.e. for how long the current state has an effect on the next steps. Since $\mathcal{W}_{res}$ is an orthogonal matrix, then $\alpha$ will express the energy decay rate of the state. $\alpha = 0$ indicates that the system is memoryless and the current state at time step $n$ depends only on the input.
\subsection{Output Layer}
The output layer transforms the reservoir state back to the macroscopic domain. Here we assume a fully connected layer, such that all the $N_{res}$ reservoir nodes are weighted and connected to $N_{ch}$ nodes of the macroscopic network. This layer practically expresses the synaptic connectivity between the different nodes of the network. It is assumed that this transformation is purely linear, taking in consideration that the overall nonlinearity of the model is dominated by the reservoir layer. The relation of the output layer is given by:
\begin{equation}
\mathbf{y}[n+1] = \mathcal{W}_{out}\mathbf{x}[n] + \mathbf{b}
\label{eq: output}
\end{equation}
where $\mathcal{W}_{out}\in \mathbb{R}^{N_{ch} 
\times N_{res}}$ is the output weight matrix, $\mathbf{b} \in \mathbb{R}^{N_{ch} \times 1}$ is a vector of biases.

Unlike $\mathcal{W}_{in}$, $\mathcal{W}_{res}$ and $\hat{\mathcal{S}}$, which are matrices with random and constrained weights, $\mathcal{W}_{out}$ has no constraints on the values of its weights, rather it is the layer which is trained with linear regression, as common in RCN approach, to obtain the desired output.

\subsection{Data Structure}
The model is, in effect, founded on a rate-coded spiking neural network. Hence the electrophysiological data required for this approach should result from multidimensional sequences of spikes. In regard to this work, electrophysiological signals were recorded by a 60 channel MEA as voltage time traces (measured around each of the electrodes). This data was preprocessed with spike and burst detection algorithms \cite{bakkum2014parameters} and exported as time traces, containing instantaneous data of spike activity counted in specific time bins. The value of the time bins is derived from the characteristic inter-burst interval (IBI) which is found in the raw data (typical value is around $4-5 ms$). In fact, this characteristic IBI value describes the typical signal propagation time between two neuronal populations. The resulting time traces are short time events ($10^2 - 10^3 ms$) and are taken from temporal network occurrences (such as network bursts or a time-windowed network response to stimulus), where activity of numerous channels is found within a specific time window. Practically, the data in this process undergoes a significant dimensionality reduction and hence no massive datasets are needed for training. 

\section{Training}
\label{sec: Training}
As in most Machine- or Deep-learning based models, the training is performed by finding the minimum value of an objective (or loss) function, while optimizing the weights between the different layers of the ANN. In particular, as was also mentioned above, the training of RCN-based models is performed only on the linear output layer, which makes the procedure computationally lighter. In fact, in the model discussed in this paper, the task is to optimize  output layer's matrix $\mathcal{W}_{out}$ and biases $\mathbf{b}$, for each input-output pair $(\mathbf{y}[n],\mathbf{y}[n+1])$ from the training data, according to \eqref{eq: input}-\eqref{eq: output}. To achieve the optimization we use the lasso regression method \cite{tibshirani1996regression}, where we find the optimal ($\mathcal{W}_{out},\mathbf{b}$), such that:
\begin{equation}
    \min_{\mathbf{w},\mathbf{b}} \Biggl(\|\mathbf{y}(\mathbf{w})-\Tilde{\mathbf{y}} \|^2  +\lambda\sum _i |w_i| \Biggr) 
    \label{eq: lasso regression}
\end{equation}
where $\Tilde{\mathbf{y}}$ is the experimental observation time-trace, $\mathbf{y}$ is the model computed time trace (\eqref{eq: input} - \eqref{eq: output}); $\mathbf{w}$ is the output matrix $\mathcal{W}_{out}$ weights; and $\lambda$ is the lasso regression parameter \cite{tibshirani1996regression}.

\section{Linearized Model and Functional Connectivity Analysis}
\label{sec: Linear model}
Let us observe the dynamics of the model. If the initial state of the reservoir is $\mathbf{x}[0] = 0$ (unexcited state), we note that, without any input $\mathbf{y}$, the time sequence of the reservoir state, $\mathbf{x}[n]$, \eqref{eq: reservoir equation non-linear}, will not change its state and as a consequence, according to \eqref{eq: input}, \eqref{eq: reservoir equation non-linear} and \eqref{eq: output}, no dynamics in the network nodes $\mathbf{y}[n]$ will be observed. Let us assume (without the loss of generality) that at a certain time-step $n = 1$ we have a small perturbation, $\mathbf{y}[1]$, such that the following holds:
\begin{equation}
\hat{\mathcal{S}}^{i,i}\sum_j{\mathcal{W}^{ij}_{in} y^j[1]} = \delta^i
\end{equation}
where $i=1,2,3\ldots N_{res}$ and $j = 1,2,3 \ldots N_{ch}$ indicate indices of each of the arrays (matrix or vector), and $\delta \ll 1$ is an arbitrary small value. In such case, for nonlinear functions that satisfy $f(\xi)\approx \xi$ for $\xi \ll 1$, we get the linear regime of \eqref{eq: reservoir equation non-linear}. If such regime is maintained in the following $k-1$ steps, from \eqref{eq: input},\eqref{eq: reservoir equation non-linear} and \eqref{eq: output} it yields that the network output at time step k+1 is:
\begin{equation}
\mathbf{y}[k+1] \approx  \sum_{n=1}^k{ \alpha^{n-1}\left\{\mathcal{T}_{n-1}\right\}\mathbf{y}[k-n+1]}
\label{eq: linear approximation}
\end{equation}
where,
\begin{equation}
\mathcal{T}_p = \mathcal{W}_{out}\hat{\mathcal{S}}[\mathcal{W}_{res}\hat{\mathcal{S}}]^p\mathcal{W}_{in}
\label{eq: transfer matrix}
\end{equation}
is a $N_{ch} \times N_{ch}$ transfer matrix of order $p$ (note that we omitted the constant bias vector $\mathbf{b}$ from \eqref{eq: output}, since it describes a constant DC offset, and as known a posteriori, its value is small).

Assuming that the regression described in Section \ref{sec: Training} (Eqn. \eqref{eq: lasso regression}) as part of the model training has achieved low training and validation error score, means that a feasible parametrization for the equations of the nonlinear model (\eqref{eq: input} - \eqref{eq: output}) has been found. It follows that the transfer matrices \eqref{eq: transfer matrix} contain the connectivity weights between the nodes in the linear regime, for different orders of interaction.

Note that by eliminating the reservoir operation, i.e., canceling the memory of previous steps, that is, taking $\alpha \to 0$ in \eqref{eq: linear approximation}, will lead to the following equation:
\begin{equation}
    \mathbf{y}[k+1] = \mathcal{T}_0 \mathbf{y}[k]
    \label{eq: zeroth order}
\end{equation}
We hence define:
\begin{equation}
    \mathcal{T}_0 = \mathcal{W}_{out}\hat{\mathcal{S}}\mathcal{W}_{in}
    \label{eq: intrinsic conn matrix}
\end{equation}
as the \textit{intrinsic connectivity matrix}, since it describes directly the weights between the network nodes for two consecutive states, regardless of the memory stored in the reservoir. Note that each component $\mathcal{T}^{i,j}_0$ shows the directed connection $j\rightarrow i$, i.e., from node $j$ at time $n$ to node $i$ at time $n+1$.
The higher order $\mathcal{T}_p$ ($p=1,2,\ldots k$) matrices contain the corrections (still in the linear regime) to the connection weights following the reservoir activation. These transfer matrices express both excitatory connections (positive values) and inhibitory connections (negative values).

\section{Conclusion}
In this paper we briefly described a computational model which decodes spatio-temporal electrophysiological data and obtains a network graph on a macroscopic scale, depicting relationships between neuronal populations. We tested this model on microelectrode array (MEA) measurements of neuronal cultures of mice cortical cells \cite{auslender2023integrated}, where the data served as the training and validation of the model.
\subsection{Retrieval of Connectivity Map}
\begin{figure}[ht]
    \centering
    \centerline{\includegraphics[width=3.0in]{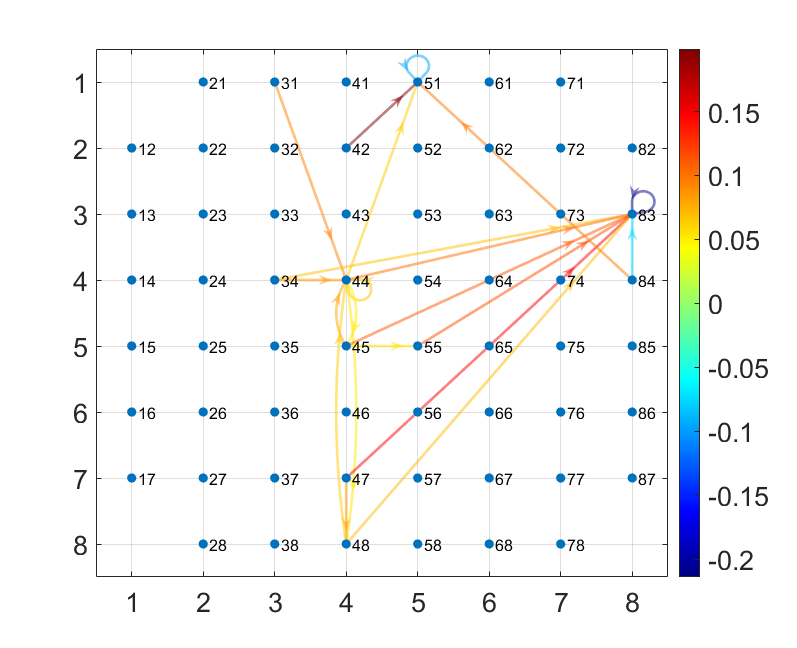}}
    \caption{An example of a connectivity map (or a graph) obtained by the ANN model discussed in this paper. The map corresponds to 60 electrode MEA layout (of $8\times 8$ matrix), where each node represents an electrode in the measurement. Each electrode samples signals from a population of neurons. The connections presented in the figure are taken from the intrinsic connectivity matrix $\mathcal{T}_0$ \eqref{eq: intrinsic conn matrix}, associated with the linearized model (Section \ref{sec: Linear model}). The model has been trained on data from microelectrode array (MEA) measurements \cite{auslender2023integrated}. Note that positive weights assigned to excitatory connections and negative to inhibitory. Threshold of $|\mathcal{T}_0| > 0.05$ was applied for visualization.}
    \label{fig:Conn Map}
\end{figure}

As shown in Section \ref{sec: Linear model}, the primary connections of the network, which do not depend on any inputs (i.e., static connections) are given by the \textit{Intrinsic connectivity matrix}, $\mathcal{T}_0$ \eqref{eq: intrinsic conn matrix}. We hence can represent these connections by a graph, where each node corresponds to a MEA electrode in the measurement. Fig. \ref{fig:Conn Map} shows the network graph obtained by the discussed model, trained on data produced by a specific neuronal culture. This graph shows the network weighted connectivity between neuronal populations sampled by the array of electrodes.

\subsection{Test and Simulation}
Given a trained model we possess the time propagation operator \eqref{Eq: Operator} (given by \eqref{eq: input}-\eqref{eq: output}), such that by giving an initial network state $\mathbf{y}[1]$ we could reproduce (or predict) the state of the network at the following time steps $\mathbf{y}[2],\mathbf{y}[3]\ldots \mathbf{y}[k]$, by propagating $\mathbf{y}[1]$ in time.

\begin{figure}[!h]
    \centering
    \includegraphics[width=3.2in]{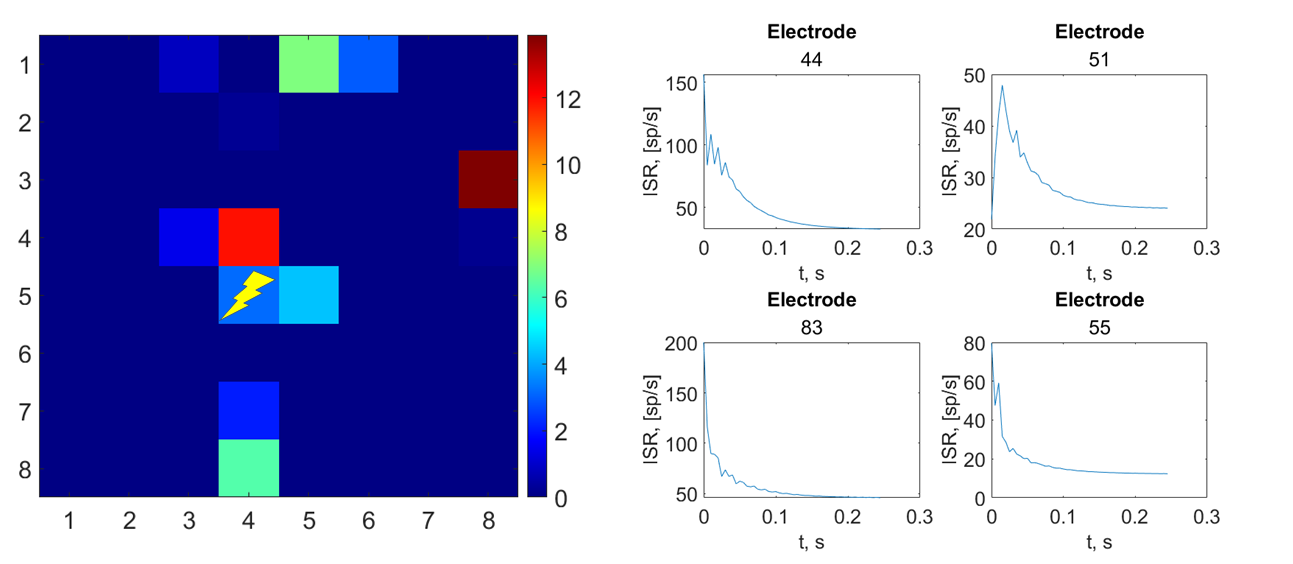}
    \caption{Simulation of network response to a local stimulus. The model has been trained on basal spontaneous activity of the culture and was tested on a response to a stimulus at electrode (node) 45 (lightning symbol on the left figure). Left: Map corresponding to $8\times 8$ MEA. Each pixel represents an electrode, the colorscale indicates the time-integrated response to the stimulus. Right: instantaneous spike rate (ISR) as a function of time (time-bin = $5ms$) of the four most responsive electrode in the network. The figures describe the time evolution of the network (response) following an initial state $\mathbf{y}[1]$ (stimulus).}
    \label{fig: response}
\end{figure}

Assuming that by training, the model has acquired the functional and structural properties of the neuronal network up to some degree of validity, it is then possible to test the response of the network to a specific input, which the model possibly has not been trained on. In such way we could simulate the response of the network, for example, to a local stimulus. In Fig. \ref{fig: response} we show an example of testing a response of the model to a local stimulus given at a specific node of the network. In this case, we trained the model on a basal spontaneous activity of 10 minutes, assuming that the functionality of the network has been learned.

Currently the model is being benchmarked with synthetic data using NEST simulator \cite{spreizer_sebastian_2022_6368024}, where we test the capacity of the model to predict the functional connectivity and the temporal response of the network. The response prediction is also being tested on experimental data of MEA recordings. A detailed description on this part of the work will be reported in a follow-up publication.

\section*{Acknowledgement}
{This work has received funding from the European Union’s Horizon 2020 research and innovation programme under the Marie Sklodowska-Curie grant agreement No 101033260 (project ISLAND) and the European Research Council (ERC) grant agreement No 788793 (project BACKUP).
}

\bibliographystyle{ieeetr}
\bibliography{PrePrint1}

\end{document}